\documentclass[a4paper,12pt]{article}
\usepackage{amssymb,amsfonts,amsmath,mathtext,cite,enumerate,float,color}
\usepackage{amssymb,amsthm,latexsym, braket, euscript,amscd, authblk}

\usepackage[russian,english]{babel}
\usepackage[utf8]{inputenc}

\title{On capacity of quantum channels generated by irreducible projective unitary representations of finite groups}

\author[1]{G.G.Amosov\thanks {This work was funded by the Ministry of Science and Higher Education of the Russian Federation (grant number 075-15-2020-788) and performed at the Steklov Mathematical Institute of the Russian Academy of Sciences.}}

\affil[1]{Steklov Mathematical Institute of Russian Academy of Sciences}

\begin{document}

\maketitle

\begin{abstract}
We study mixed unitary quantum channels generated by irreducible projective unitary representations of finite groups. Under some assumptions on the probability distribution determining a mixture the classical capacity of the channel is found. Examples illustrating the techniques are given.
\end{abstract}

\section{Introduction}

The quantum coding theorem \cite {Holevo, SW} gives rise to the estimation from above for a number of states $N$ that can be asymptotically accurately transmitted through $m$ copies of the quantum channel $\Phi $ of the form
$$
N\sim e^{mC({\Phi })},\ m\to +\infty ,
$$
where $C(\Phi )$ is said to be a classical capacity of $\Phi $. 
The capacity $C(\Phi )$ can be calculated by the formula
$$
C(\Phi )=\lim \limits _{m\to \infty }\frac {C_1(\Phi ^{\otimes m})}{m},
$$
where $C_1 (\Phi )$ is  a one-shot capacity determined by the Holevo upper bound as follows
$$
C_1 (\Phi )=\sup \limits _{\pi _j,\rho _j\in {\mathfrak S}(H)}(S(\sum \limits _j\pi _j\Phi (\rho _j))-\sum \limits _j\pi _jS(\Phi (\rho _j))).
$$
Here  $S(\rho )=-Tr(\rho \log \rho )$ is the von Neumann entropy of quantum states $\rho _j$ from the set of all states $\mathfrak {S}(H)$ in a Hilbert space $H$ and the supremum is taken over all probability distributions $(\pi _j)$.

The main feature of quantum theory is the presence of entangled states in $\mathfrak {S}(H^{\otimes m})$, that is, such states that are not factorized with respect to the tensor product $H^{\otimes m}$. The presence of entangled states leads to the fact that taking the supremum in the definition of $C_1(\Phi ^{\otimes m})$ becomes a very difficult mathematical problem. If the equality of the form
\begin{equation}\label{Hast}
C_1(\Phi ^{\otimes m})=mC_1(\Phi )
\end{equation}
were true, it would mean that the use of entangled states does not give an advantage in encoding. In the last case, the capacity of $\Phi $ coincides with the one-shot capacity
\begin{equation}\label{key}
C(\Phi )=C_1(\Phi ).
\end{equation}
Nevertheless it is known that (\ref {key}) doesn't take place in general \cite{Hast}.

It is natural to start calculating the capacity of a channel with the case of mixed unitary channels having the form
\begin{equation}\label{KING}
\Phi (\rho )=\sum \limits _{j}\pi _jU_j\rho U_j^*,
\end{equation}
where $\rho \in {\mathfrak S}(H)$, $(\pi _j)$ is a probability distribution and $(U_j)$ are unitary operators in $H$. Operators $U_j$ simulate errors that may occur during the transmission of information. Since the content part of the problem concerns the study of possible encodings in the tensor product $H^{\otimes m}$, it turns out that only error operators acting locally on the qudit $H$ forming a composite system $H^{\otimes m}$ are allowed. For the qubit case $dimH=2$ quantum channels of the form (\ref {KING}) were studied in \cite{King1}. The property (\ref {key}) was proved and the capacity of channels were found. Here unitary operators in (\ref {KING}) may mean bit flip or sign (phase) flip errors. 
In \cite{King2} the same problem was solved for the quantum depolarizing channel having the form (\ref {KING}) in any dimension. Both the cases considered relate to channels of a narrower type than (\ref {KING}). More precisely, these channels have the form
\begin{equation}\label{KING2}
\Phi (\rho )=\sum \limits _{g\in G}\pi _gU_g\rho U_g^*,
\end{equation}
where $G\ni g\to U_g$ is an irreducible projective unitary representation of a finite group $G$ and $(\pi _g)$ is a probability distribution on $G$. Moreover, as we will show and discuss it appeared that $(U_g,\ g\in G)$ generate the group ${\mathcal U}_G$ containing a maximum Abelian normal subgroup. It should be noted that the hypothetical counterexample of \cite{Hast} has the form (\ref {KING}) but we suppose that it doesn't belong to the class (\ref{KING2}).

Since the unital channel (\ref {KING2}) is covariant with respect to the irreducible projective unitary representation $g\to U_g$ we obtain that \cite {Holevo2}
\begin{equation}\label{okonc}
C_1(\Phi )=\log n-\inf \limits _{\rho \in \mathfrak {S}(H)}S(\Phi (\rho )).
\end{equation}  
If the property
\begin{equation}\label{Hast}
\inf \limits _{\rho \in \mathfrak {S}(H^{\otimes ^m})}S(\Phi ^{\otimes ^m}(\rho ))=m\inf \limits _{\rho \in \mathfrak {S}(H)}S(\Phi (\rho ))
\end{equation} 
known as the weak additivity for output entropy holds true the classical capacity of $\Phi $ can be found by means of a simple formula
\begin{equation}\label{Chingiz}
C(\Phi )=C_1(\Phi )=\log n-\inf \limits _{\rho \in \mathfrak {S}(H)}S(\Phi (\rho )).
\end{equation}
At the moment, (\ref {Chingiz}) is proved for only few partial cases \cite {King1, King2, Amo}. 

Recently, there has been renewed interest in computing the capacity of quantum channels, related to the application of the majorization procedure \cite{Hol, Hol2, urReh, Siu, Amo}. The idea of applying the majorization procedure \cite {Kar} for channels in a finite dimensional Hilbert space $H,\ dimH=n<+\infty,$ is proposed in \cite{urReh} and based upon the division of elements of the probability distribution $(\pi _g)$ from (\ref {KING2}) into subsets, the sums of the elements of which are placed in descending order. This allows us to obtain estimates of eigenvalues for $\Phi (\rho )$. However, in order for the estimates to be achievable, some additional conditions must be met. In \cite{Siu} it was shown that if the conditions of achievability for \cite{urReh} are fulfilled for $\Phi$ it would be not true for $\Phi ^{\otimes m}$. We will show that this complexity can be overcome.

This work is a continuation of \cite {Amo}, where the classical capacity was calculated for the Weyl channels that are perturbations of quantum-classical (q-c) channels. Here we justify the results of \cite {Amo} and develope the techniques of \cite{Amo} to the channels generated by irreducible projective unitary representations of discrete groups. We use the ideas that originate from \cite{Amo+, A1, A2, A3}. Applying the techniques of the group theory was inspired by \cite {Zhda1, Zhda2}. These papers in turn were encouraged by \cite{Shirokov, Shirokov2, Shirokov3}.

This paper is organized as follows. We start with Preliminaries containing basic notations used in the text. The next two section are devoted  to basic results from the theory of majorization and the theory of representations for finite groups. Our main result concerning the calculation of one-shot and classical capacities for mixed unitary channels is placed in Section 5. The proof is given in Section 6. Section 7 is devoted to the construction of examples. The paper is completed by concluding remarks. 

\section{Preliminaries}

Throughout this paper we use the following standard notations of quantum information theory \cite{Holevo2} and the group theory
\cite{Hall, Ham}.

$H$ is the Hilbert space with the dimension $dimH=n$ and  the fixed orthonormal basis $\ket {j},\ j\in {\mathbb Z}_n$;

$B(H)$ is the algebra of all bounded operators in $H$;

${\rm I}={\rm I}_H$ is the identity operator in $H$;

$\mathfrak {S}(H)$ the convex set of all states (positive unit trace operators) in $H$;

$\mathcal {U}(H)$ the group of all unitary operators in $H$;

Given $\rho \in \mathfrak {S}(H)$ we denote $S(\rho )=-Tr(\rho \log \rho )$ the von Neumann entropy of $\rho $;

The linear map $\Phi :B(H)\to B(H)$ is said to be a quantum channel if it is completely positive and trace preserving such that $\Phi :\mathfrak {S}(H)\to \mathfrak {S}(H)$;

The channel $\Phi $ is said to be a quantum-classical (q-c) channel if $\Phi (\rho )$ belongs to the maximum Abelian algebra ${\mathcal A}\subset B(H)$ for all $\rho \in {\mathfrak S}(H)$;

${\mathbb Z}_n$ is the cyclic group with elements $\{0,\dots ,n-1\}$;

$S_n$ is the symmetric group consisting of all permutations on the set of indexes $\{0,\dots ,n-1\}$;

$S_n$ is the set of all permutations of indexes $j\in \{0,1,2,\dots ,n-1\}$ identifying with elements $j\in {\mathbb Z}_n$;

A subgroup $T$ of the group $G$ is said to be normal if $gtg^{-1}\in T$ for all $g\in G,\ t\in T$. We denote elements of the quotient $G/T$ by $[g]$.

\section{Majorization}

Here we give the basic concept of the majorization theory \cite{Kar} following to \cite{Bha}.

Given a probability distribution $\mu =\{\mu _l,\ l\in {\mathbb Z}_n\}$ denote $\mu ^{\downarrow}$ the same distribution rearranged in the descending order.
For two probability distributions $\mu =\{\mu _l,\ l\in {\mathbb Z}_n\}$ and $\nu =\{\nu _l,\ l\in {\mathbb Z}_n\}$ we say that
$\mu $ majorizes $\nu $ and write \cite{Bha}
$$
\nu ^{\downarrow}\prec \mu ^{\downarrow}
$$
iff
$$
\sum \limits _{l=0}^r\nu _l^{\downarrow} \le \sum \limits _{l=0}^{r}\mu _l^{\downarrow},\ 0\le r<n. 
$$
The famous characteristic of majorization is placed below.

{\bf Theorem II.3.1 \cite {Kar, Bha}.} {\it Suppose that $\mu =\mu ^{\downarrow}$ and $\nu =\nu ^{\downarrow}$. The following two conditions are equivalent:

(i)  $\nu \prec \mu $.

(ii) $\sum \limits _j \phi (\nu _j)\le \sum \limits _j\phi (\mu _j)$ for all convex functions $\phi $ from $\mathbb R$ to $\mathbb R$.
}

Applying Theorem given above we obtain that the following inequalities hold true
\begin{equation}\label{Bha1}
\sum \limits _{l\in {\mathbb Z}_n}\nu _l^p\le \sum \limits _{l\in {\mathbb Z}_n}\mu _l^p,\ p\ge 1,
\end{equation}
and
\begin{equation}\label{Bha2}
-\sum \limits _{l\in {\mathbb Z}_n}\nu _l\log\nu _l\ge -\sum \limits _{l\in {\mathbb Z}_n}\mu _l\log \mu _l.
\end{equation}
The inequality (\ref {Bha1}) concerns the multiplicativity property for $l_p$-norms \cite{AH}. We especially interested in the inequality (\ref {Bha2}) allowing to calculate a classical capacity of the channel under the conditions we consider.

\section{Projective unitary representations of finite groups}

Here we put the standard facts of the representation theory \cite{Hall, Ham}.

Let $G$ be a finite group. 
The map $G\ni g\to U_g\in {\mathcal U}(H)$ is said to be a projective unitary representation of $G$ if
\begin{equation}\label{1}
U_{g}U_h=\omega (g,h)U_{gh},
\end{equation}
where $\omega (g,h)$ is an unimodular function  on $G\times G$. 
In order for (\ref {1})  to be correct, the following condition must be met 
\begin{equation}\label{2}
\omega (g,h)\omega (gh,r)=\omega (g,hr)\omega (h,r),\ g,h,r\in G.
\end{equation}
Function $\omega (g,h)$ satisfying the property (\ref {2}) is said to be $2$-cocycle of $G$.
Two representation $g\to U_g$ and $g\to V_g$ are said to be equivalent if there exists the unimodular function $c(g)$
such that
\begin{equation}\label{3}
V_g=c(g)U_g,\ g\in G.
\end{equation}
It is straightforward to check that 
$$
V_gV_h=\tilde \omega (g,h)V_{gh},
$$
with
$$
\tilde \omega (g,h)=\frac {c(g)c(h)}{c(gh)}\omega (g,h).
$$
Given a representation $g\to U_g$ of the group $G,\ |G|=N$ there exists the equivalent representation
with the function $\omega (g,h)$ satisfying \cite{Ham}
\begin{equation}\label{root}
\omega (g,h)^N=1,\ g,h\in G.
\end{equation}
The $2$-cocycles $\omega (g,h)$ form the Abelian group with respect to a pointwise multiplication. 
This group is said to be the Schur multiplier. Adding to the unitary operators $\{U_g,\ g\in G\}$ the Schur multiplier we obtain the unitary group acting in the Hilbert space $H$ \cite {Ham}. We denote ${\mathcal U}_G$ this group.

\section{The main result}

Let $T\triangleleft G$ be a normal Abelian subgroup of $G$. 
Consider a unitary representation of $t\to U_t$ of the group $T$ in the Hilbert space $H,\ dimH=n$. Without a sake of generality we can suppose that the operators $U_t,\ t\in T,$ are diagonal in the basis $(\ket {j})$.
Let $g\to U_g$ be a projective unitary representation  of $G$ in a Hilbert space $H$ extending the representation $t\to U_t$ of $T$.

Let us define a mixed unitary quantum channel by the formula
\begin{equation}\label{REV}
\Phi (\rho )=\sum \limits _{g\in G}\pi _gU_g\rho U_g^*,\ \rho \in \mathfrak {S}(H),
\end{equation}
where $\pi =(\pi _g,\ g\in G)$ is a probability distribution on $G$.

Together with the probability distribution $(\pi _g,\ g\in G)$ determining the channel (\ref {REV}) let us consider the probability distribution $p_{[g]}$ on the quotient $G/T$ defined by the formula
\begin{equation}\label{probab}
p_{[g]}=\sum \limits _{t\in T}\pi _{gt}.
\end{equation}
Denote $p=(p_j,\ j\in {\mathbb Z}_n)$ values (\ref {probab}) placed in descending order such that
$$
p_0\ge p_1\ge \dots \ge p_{n-1}
$$
Let $[g_j]$ be the element of $G/T$ such that $p_{[g_j]}=p_j$. Let us define the following condition
\begin{equation}\label{cond}
j>k \Rightarrow \pi _{g_jt}\le \pi _{g_ks}
\end{equation}
for all $s,t\in T$ and $j,k\in {\mathbb Z}_n$.

{\bf Theorem.} {\it Let $G$ and $T$ be a finite group and its normal Abelian subgroup with the property $|G/T|=n$. Suppose that $g\to U_g$
is an irreducible projective unitary representation of $G$ in a Hilbert space $H,\ dimH=n$, and the restriction $t\to U_t,\ t\in T$ is unitary representation. Then, if the condition (\ref {cond}) is satisfied the one-shot and classical capacities of (\ref {REV}) are given by the formula
$$
C(\Phi )=C_1(\Phi )=\log n+\sum \limits _{j\in {\mathbb Z}_n}p_j\log p_j.
$$
}

The rest of the text is devoted to clarifying the condition (\ref {cond}), proving Theorem, and giving several examples of how to apply Theorem.

\section{Quantum channels generated by projective unitary representations}

 The following proposition shows how the projective unitary representation of the group satisfying the conditions of Theorem is arranged.

{\bf Proposition 1.} {\it Let $T$ be a normal Abelian subgroup of $G$ and $g\to U_g$ is an irreducible projective unitary representation of $G$ in $H$ extending the unitary representation $t\to U_t$ of $T$. Suppose that $|G/T|=n$, then there is the action $\alpha_{[g]}$ of $G/T$ consisting of permutations on ${\mathbb Z}_n$ considered as a set such that
\begin{equation}\label{CO}
U_{g}\ket {j}\bra {j}U_{g}^*=\ket {\alpha_{[g]}(j)}\bra {\alpha_{[g]}(j)},
\end{equation}
$g\in G,\ j\in {\mathbb Z}_n$.
}

Proof.

Since the representation is irreducible we get
\begin{equation}\label{FORM}
\frac {1}{|G|}\sum \limits _{g\in G}U_g\ket {f}\bra {f}U_g^*=\frac {1}{n}{\rm I}
\end{equation}
for any unit vector $f\in H$. Projections $\ket {j}\bra {j}$ belong to the algebra of fixed elements with respect to the actions of 
$B(H)\ni x\to U_txU_t^*,\ t\in T$. Substituting $\ket {f}=\ket {j}$ to (\ref {FORM}) we obtain
\begin{equation}\label{FORM2}
\sum \limits _{[g]\in G/T}U_g\ket {j}\bra {j}U_g^*={\rm I}
\end{equation}
because $|G/T|=n$ by the condition.
The sum of projections in (\ref {FORM2}) is equal to the identity operator only if they are pairwise orthogonal.

$\Box $

The next statement explains how the majorization works.

{\bf Proposition 2.} {\it Suppose that the conditions of Proposition 1 are satisfied and the condition (\ref{cond})
holds true.
Then, given a unit vector $f\in H$
the eigenvalues $\lambda =(\lambda _m)$ of the operator $\Phi (\ket {f}\bra {f})$ satisfy the relation
$$
\lambda ^{\downarrow}\prec p.
$$
}

Proof.

Since the representation $g\to U_g$ is irreducible (\ref {FORM}) is fulfilled. Because $|G/T|=n$ we can rewrite (\ref {FORM}) as follows
\begin{equation}\label{FORM3}
\sum \limits _{[g_{j}]\in G/T,t\in T}|\braket {e,U_{g_jt}f}|^2=|T|
\end{equation}
for any unit vector $e\in H$.
Let $(e_m)$ be the eigenvectors corresponding to the eigenvalues $\lambda ^{\downarrow}=(\lambda _m^{\downarrow})$. It follows from (\ref {FORM3}) that
$$
\sum \limits _{m=0}^k\lambda _m^{\downarrow}=\sum \limits _{m=0}^k\braket {e_m,\Phi (\ket {f}\bra {f})e_m}=
$$
\begin{equation}\label{MAGIA}
\sum \limits _{m=0}^k\sum \limits _{[g_{j}]\in G/T,t\in T}\pi _{g_jt}|\braket {e_m,U_{[g_j]t}f}|^2\le \sum \limits _{m=0}^kp_m,\ 0\le k\le n-1,
\end{equation}
due to each sum in (\ref {probab}) consists of $|T|$ elements and (\ref {cond}).

$\Box $

Consider the partial case of probability distribution satisfying (\ref {cond})
$$
\pi _{g_jt}=\frac {1}{n}p_j,\ j\in {\mathbb Z}_n
$$
and the corresponding channel
\begin{equation}\label{partial}
\Phi _0(\rho )=\frac {1}{n}\sum \limits _{j\in {\mathbb Z}_n,t\in T}p_jU_{g_jt}\rho U_{g_jt}^*,\ \rho \in {\mathfrak S}(H).
\end{equation}

{\bf Proposition 3.} {\it The channel (\ref {partial}) is the q-c channel with the image belonging to the maximum Abelian subalgebra $\mathcal A$ generated by $\ket {j}\bra {j},\ j\in {\mathbb Z}_n$.}

Proof.

The conditional expectation $\Theta $ on the algebra $\mathcal A$ is given by the formula
$$
\Theta (x)=\frac {1}{|T|}\sum \limits _{t\in T}U_txU_t^*,\ x\in B(H).
$$ 
It is straightforward to check that
$$
\Theta \circ \Phi _0=\Phi _0.
$$

$\Box $

The quantum channel $\Phi $ possessing the property (\ref {cond}) is naturally to call a perturbation of q-c channel (\ref {partial}).

Now we are ready to proceed to the study of the properties of channels for tensor products.
Given a probability distribution $q$ on  ${\mathbb Z}_n$ denote $q^{\oplus N}$ the probability distribution $(q_0,q_1,\dots ,q_{n-1},0,\dots ,0)$, where $(n-1)N$ zeros are added to the distribution $q$.

{\bf Proposition 4.} {\it Fix the probability distribution $q$ on ${\mathbb Z}_{n}$. Suppose that the conditions of Proposition 1 are satisfied and
$$
\lambda ^{\downarrow}\prec q,
$$
for the eigenvalues $\lambda $ of the state $\Phi (\ket {e}\bra {e})$ under any choice of a unit vector $e\in H$.
Consider $Id \otimes \Phi :\mathfrak {S}(K\otimes H)\to \mathfrak {S}(K\otimes H)$.  Then, given a unit vector $f\in K\otimes H$ the eigenvalues $\mu =(\mu _j)$ of the state $(Id \otimes \Phi )(\ket {f}\bra {f})$ satisfy the relation
$$
\mu ^{\downarrow}\prec q^{\oplus N},
$$
where $N=dim K$.
}

Proof.

Consider the subspace ${\mathcal L}\subset K\otimes H$ generated by vectors $(I\otimes U_g)f,\ g\in G$. Since the representation $g\to U_g$ is irreducible and $|G|=n^2$ the unitary operators $(U_g,\ g\in G)$ are linearly independent.
Hence $dim{\mathcal L}=n$ and $G\ni g\to ({\rm I}\otimes U_g)|_{\mathcal L}$ is the irreducible projective unitary representation of $G$ in $\mathcal L$. Moreover, $U_g\to (I\otimes U_g)|_{\mathcal L}$ defines a unitary representation of the group ${\mathcal U}_G$. Since the representation $U_g\to I\otimes U_g$ of the group ${\mathcal U}_G$ can be represented as the orthogonal sum of $n$ irreducible representations each of which is equivalent to ${\mathcal U}_G$ and the resolution of a reducible representation in the sum of orthogonal irreducible representations is unique up to the permutation of terms \cite{Hol} we get that the group generated by $(I\otimes U_g)|_{\mathcal L},\ g\in G,$ is unitarily equivalent to ${\mathcal U}_G$. Hence, the channel
$Id\otimes \Phi:\mathfrak {S}({\mathcal L})\to \mathfrak {S}({\mathcal L})$ has the same characteristics as $\Phi $.  

$\Box $

{\bf Proposition 5.} {\it Let $\Phi :\mathfrak {S}(H)\to \mathfrak {S}(H)$ and $\Omega :\mathfrak {S}(K)\to \mathfrak {S}(K)$ be two channels of the form (\ref {REV}), $\Phi $ satisfies the conditions of Propositions 1, 2 and there exists the probability distribution $q=(q_j)$ such that the eigenvalues $\lambda =(\lambda _j)$ of $\Omega (\ket {e}\bra {e})$ obeys the condition
$$
\lambda ^{\downarrow}\prec q
$$
for any choice of unit vector $e\in K$. Then, the eigenvalues $\mu =(\mu _j)$ of $(\Phi \otimes \Omega )(\ket {f}\bra {f})$
satisfies
$$
\mu ^{\downarrow}\prec (pq)^{\downarrow}
$$
for any choice of unit vector $f\in H\otimes K$.
}

Proof.

Given a unit vector $f\in H\otimes K$ denote $\lambda ^f=(\lambda _j^f)$ the eigenvalues of $(Id\otimes \Omega )(\ket {f}\bra {f})$ arranged in the descending order. It follows from Proposition 4 that
$$
\lambda ^f\prec q^{\oplus dimH}.
$$
Denote $\Pi =(\Pi _j)$ the probability distribution $(p_kq_m)$ arranged in the descending order,
$$
\Pi _0\ge \Pi _1\ge \dots \ge \Pi_{ndimK-1}.
$$
Given a unit vector $f\in H\otimes K$ let us consider 
$$
(\Phi \otimes \Omega )(\ket {f}\bra {f})=\sum \limits _{[g_j]\in G/T,t\in T}\pi _{g_jt}(U_{g_jt}\otimes I_K)((Id\otimes \Omega )(\ket {f}\bra {f}))(U_{g_jt}^*\otimes I_K).
$$
Notice that
\begin{equation}\label{proof1}
\sum \limits _{[g_j]\in G/T,t\in T}(U_{g_jt}\otimes I_K)((Id\otimes \Omega) (\ket {f}\bra {f}))(U_{g_jt}^*\otimes I_K)=n{\rm I}_H\otimes \Omega (Tr_H(\ket {f}\bra {f})).
\end{equation}
The identity (\ref {proof1}) implies that
\begin{equation}\label{end1}
\sum \limits _{[g_j]\in G/T,t\in T}\braket {h_m,(U_{g_jt}\otimes I_K)(Id\otimes \Omega )(\ket {f}\bra {f})(U_{g_jt}^*\otimes I_K)h_m}\le n.
\end{equation}
\noindent
It follows from Proposition 4 that
\begin{equation}\label{end2}
\sum \limits _{m=0}^{k-1}\braket {h_m,(U_{g_jt}\otimes I_K)((Id\otimes \Omega) (\ket {f}\bra {f}))(U_{g_jt}^*\otimes I_K)h_m}\le \sum \limits _{m=0}^{k-1}q_{m}^{\oplus dimH}
\end{equation}
for any choice of $k$ pairwise orthogonal unit vectors $h_m\in H\otimes K$.
Thus, we obtain
$$
\sum \limits _{m=0}^{k-1}\braket {h_m,(\Phi \otimes \Omega )(\ket {f}\bra {f})h_m}=
$$
$$
\sum \limits _{m=0}^{k-1}\sum \limits _{[g_j]\in G/T,t\in T}\pi _{g_jt}(U_{g_jt}\otimes I_K)((Id\otimes \Omega )(\ket {f}\bra {f}))(U_{g_jt}^*\otimes I_K)\le
$$
$$
\sum \limits _{m=0}^{k-1}\Pi _m,\ 0\le k\le ndimK,
$$
where we have combined (\ref {end1}), the majorization condition (\ref {cond}) and (\ref {end2}).

$\Box $

Proof of Theorem.

The value for one-shot capacity is immediately follows from Propositions 1, 2 and (\ref {Bha2}). 
Suppose that the conditions of Proposition 5 are satisfied for some channel $\Omega $. Hence,
given a unit vector $f\in H\otimes K$ for the eigenvalues $\mu =(\mu _j)$ of $(\Phi \otimes \Omega )(\ket {f}\bra {f})$
we obtain
\begin{equation}\label{end}
\mu ^{\downarrow}\prec (pq)^{\downarrow}
\end{equation}
due to Proposition 5. Put $\Omega =\Phi $, then the channel $\Phi \otimes \Phi $ satisfies (\ref {end}). Thus, we can apply the same procedure for $\Phi $ and $\Omega =\Phi \otimes \Phi $. Denote $\Pi $ the probability distribution
$(\prod _{j=1}^Np_{s_j})$. Acting consistently, we get for the eigenvalues $\lambda =(\lambda _j)$ of $\Phi ^{\otimes N}(\ket {f}\bra {f})$ with an arbitrary unit vector $f\in H^{\otimes N}$
$$
\mu ^{\downarrow }\prec \Pi ^{\downarrow}.
$$
Hence, the minimal output entropy
$$
\inf \limits _{\rho \in {\mathfrak S}(H^{\otimes N})}(\Phi (\rho ))=-N\sum \limits _{j\in {\mathbb Z}_n}p_j\log p_j
$$
due to (\ref {Bha2}). Now the result follows from (\ref {okonc}).

$\Box$

\section{Construction of representations.}

In this section we give several examples of irreducible projective unitary representations of Abelian and non-Abelian groups for which Theorem can be applied. Surely there are many other examples, see e.g. \cite{Jan}.

Let $T={\mathbb Z}_n$ and $\mathcal S$ be a subgroup  of the symmetric group $S_n$. Notice that any finite group is isomorphic to a subgroup of $S_n$ for some $n$ by the Caley theorem. There is a natural action of $\mathcal S$ on $T$ considered as a set of $n$ elements which is determined by permutations as
$$
\alpha _s(t)=s(t),\ s\in {\mathcal S},\ t\in T.
$$

Fix the set $E =(\chi _0,\dots ,\chi _{n-1})$ consisting of $n$ characters of $T$,
$$
\chi _j(tt')=\chi _j(t)\chi _j(t'),\ t,t'\in T,
$$
giving rise to $\chi _j(t)^{n}=1,\ j\in {\mathbb Z}_n$. Characters from $E$ can coincide. Let us define unitary operators $W_t\in {\mathcal U}(H)$ by the formula
\begin{equation}\label{W}
W_t\ket {j}=\chi _j(t)\ket {j},\ j\in {\mathbb Z}_n,\ t\in T.
\end{equation}
Consider the representation of $\mathcal S$ in $H$ determined as follows
\begin{equation}\label{V}
V_s\ket {j}=\ket {s(j)},\ j\in {\mathbb Z}_n,\ s\in {\mathcal S}.
\end{equation}
The next statement immediately follows from Theorem.

{\bf Corollary.} {\it Suppose that (\ref {W}) and (\ref {V}) determine the irreducible projective unitary representation of the group $G={\mathcal S}\times {\mathbb Z}_n$. Then, if the condition (\ref {cond}) is satisfied the classical capacity of (\ref {REV}) is given by the formula
$$
C(\Phi)=C_1(\Phi )=\log n+\sum \limits _{j\in {\mathbb Z}_n}p_j\log p_j.
$$
}

{\bf Example 1.} {\it Qubit channels.}

Here $G={\mathbb Z}_2\times {\mathbb Z}_2$ and ${\mathcal U}_G$ is the group generated by the Pauli matrices. Let us define the characters as follows $\chi _0(0)=\chi _0(1)=\chi _1(0)=1,\ \chi _1(1)=-1$ and the action $s(0)=1$. 
Any unital qubit channel $\Phi $ is determined by the parameters $(\lambda _x,\lambda _y,\lambda _z)$ giving the action on the Pauli matrices \cite{King1}
\begin{equation}\label{pi}
\Phi (\sigma _x)=\lambda _{x}\sigma _x,\ \Phi (\sigma _y)=\lambda _{y}\sigma _y,\ \Phi (\sigma _z)=\lambda _{z}\sigma _z.
\end{equation}
Four numbers including in the probability distribution $\pi $ determining a channel $\Phi$ of the form (\ref {REV}) can be arranged in the descending order. If maximal one is related to the action $\sigma _a\cdot \sigma _a$, $a\in \{x,y,z\}$ we can consider $\tilde \Phi (\cdot )=\sigma _a\Phi (\cdot )\sigma _a$. Now the maximal value in the distribution $\pi $ corresponds to the identical action. There are three possibility for determining a unitary representation of the Abelian group ${\mathbb Z}_2$. These are $0\to {\rm I},\ 1\to \sigma _a,\ a\in \{x,y,z\}$. For any of them we get  the probabilities in the mixture corresponding to (\ref {pi}) as follows
$$
\pi _{00}+\pi _{01}-\pi _{10}-\pi _{11}=\lambda _x,
$$
$$
\pi _{00}+\pi _{10}-\pi _{01}-\pi _{11}=\lambda _y,
$$
$$
\pi _{00}+\pi _{11}-\pi _{01}-\pi _{10}=\lambda _z.
$$
Taking the majorization procedure over the set $\{\pi _{00},\pi _{01},\pi _{10},\pi _{11}\}$ we obtain
$$
C(\Phi)=C_1(\Phi )=\log 2+\frac {1+\lambda _{max}}{2}\log \frac {1+\lambda _{max}}{2}+\frac {1-\lambda _{max}}{2}\log \frac {1-\lambda _{max}}{2},
$$
where
$$
\lambda _{max}=\max (|\lambda _x|,|\lambda _y|,|\lambda _z|).
$$

{\bf Example 2.} {\it $G={\mathbb Z}_n\times {\mathbb Z}_n$ and ${\mathcal U}_G$ is the Heisenberg-Weyl group.}

Let ${\mathcal S}=T={\mathbb Z}_n$ and the set $E$ consists of pairwise different characters $\chi_k,\ k\in {\mathbb Z}_n$ 
$$
\chi _k(j)=e^{\frac {2\pi kj}{n}},\ k,j\in {\mathbb Z}_n. 
$$
Suppose that
${\mathbb Z}_n$ acts on ${\mathbb Z}_n$ as shifts modulo $n$. Then, 
the unitary operators (\ref {W}) and (\ref {V}) generates the Heisenberg-Weyl group. 
Here we give the example constructed in \cite{Amo} for $n=3$. 
Put
$$
\pi _{00}=\frac {1}{4},\ \pi _{10}=\frac {1}{8},\ \pi _{20}=\frac {1}{8},
$$
\begin{equation}\label{example}
\pi _{01}=\frac {1}{8},\ \pi _{11}=\frac {1}{8},\ \pi _{21}=\frac {1}{12},
\end{equation}
$$
\pi _{02}=\frac {1}{12},\ \pi _{12}=\frac {1}{24},\ \pi _{22}=\frac {1}{24}
$$
and consider the channel
$$
\Phi (\rho )=\sum \limits _{k,m\in {\mathbb Z}_3}\pi _{km}V^kW^m\rho W^{*m}V^{*k},\ \rho \in \mathfrak {S}(H).
$$
Denote
$$
p_k=\sum \limits _{m\in {\mathbb Z}_3}\pi _{km},\ k\in {\mathbb Z}_3,
$$
then
$$
p_0=\frac {1}{2},\ p_1=\frac {1}{3},\ p_2=\frac {1}{6}.
$$
It follows from Theorem
\begin{equation}\label{example3}
C(\Phi)=C_1(\Phi )=\log (3)-\frac {1}{2}\log (2)-\frac {1}{3}\log (3)-\frac {1}{6}\log (6).
\end{equation}

{\bf Example 3.} {\it $G={\mathbb K}_4\times {\mathbb Z}_4$.}

Let ${\mathcal S}={\mathbb K}_4$ be the Klein group generated by a unit $e$ and three elements $x,y,z$ satisfying the relations
$$
x^2=y^2=z^2=e,\ xy=yx=z.
$$
Let us define the action $s$ of ${\mathbb K}_4$ on ${\mathbb Z}_4$ by the rule
$$
s_x(0)=1,\ s_x(1)=0,\ s_x(2)=3,\ s_x(3)=2,
$$
$$
s_y(0)=2,\ s_y(2)=0,\ s_y(1)=3,\ s_y(3)=1,
$$
$$
s_z(0)=3,\ s_z(3)=0,\ s_z(1)=2,\ s_z(2)=1.
$$
Put $\chi _0(1)=\chi _1(1)=\chi _0(2)=\chi _2(2)=1,\ \chi _2(1)=\chi _3(1)=\chi _1(2)=\chi _3(2)=-1$. 
Now we can define a projective unitary representation of ${\mathbb K}_4\times {\mathbb Z}_4$ by the formula
\begin{equation}\label{UNI}
(h,k)\to V_{s_h}W_k,\ h\in {\mathbb K}_4,\ k\in {\mathbb Z}_4
\end{equation}
with $V_{s_h}$ and $W_k$ determined by (\ref {V}) and (\ref {W}).

{\bf Proposition 6.} {\it The representation (\ref {UNI}) is irreducible.}

Proof.

The commutant of $V_{s_h},\ h\in {\mathbb K}_4$ consists of matrices
\begin{equation}\label{mat}
\begin{pmatrix}
a & b & 0 & 0\\
b & a & 0 & 0\\
0 & 0 & a & b \\
0 & 0 & b & a 
\end{pmatrix},
\end{equation}
where $a,b \in {\mathbb C}$.
The matrices of the form (\ref {mat}) commuting with $W_k,\ k\in {\mathbb Z}_4$ are multiplies of identity.

$\Box $

Proposition 6 implies that the mixed unitary channel (\ref {REV}) with $G={\mathbb K}_4\times {\mathbb Z}_4$ for which the majorization condition is satisfied has the capacities given by Corollary. 

{\bf Example 4.} {\it $G={\mathbb D}_n\times {\mathbb Z}_{2n}$.}

Dihedral group ${\mathbb D}_n$ is generated by unit $e$ and two elements $x,y$ satisfying the relation
$$
x^n=y^2=e,\ xy=yx^{-1}.
$$
While the groups ${\mathbb D}_1={\mathbb Z}_2$ and ${\mathbb D}_2={\mathbb K}_4$ are Abelian, all other groups ${\mathbb D}_n,\ n>2,$ are non-Abelian. Let us define the action of ${\mathbb D}_n$ on ${\mathbb Z}_{2n}$ as follows
\begin{equation}\label{ACT}
s_x(j)=j+2\ mod\ 2n,\ 0\le j\le 2n,\ s_y(2j)=2j+1,\ 0\le j\le n-1.
\end{equation}
Put $\chi _{2j}(1)=e^{i\frac {2\pi j}{n}},\ \chi _{2j+1}(1)=-e^{i\frac {2\pi j}{n}},\ j\in {\mathbb Z}_{n}$. Define the irreducible projective unitary representation of $G={\mathbb D}_n\times {\mathbb Z}_{2n}$ by the formula
\begin{equation}\label{non}
(h,k)\to V_{s_h}W_k,\ h\in {\mathbb D}_n,\ k\in {\mathbb Z}_{2n}
\end{equation}
with $V_{s_h}$ and $W_k=W_1^k$ determined by (\ref {V}) and (\ref {W}). 

{\bf Proposition 7.} {\it The representation $G$ defined by (\ref {non}) is irreducible.}

Proof.

The Abelian group $W_k,\ k\in {\mathbb Z}_{2n}$ generates the maximal Abelian algebra $\mathcal A$ consisting of all operators that are diagonal in the basis $\ket {j}$ because all eigenvalues of $W_1$ are pairwise different. Applying to $x\in \mathcal A$ the unitary operators $V_{s_h}$ we see that $V_{s_h}xV_{s_h}^*=x$ for all $h\in {\mathbb D}_n$ only if $x=c{\rm I}$ for some $c\in {\mathbb C}$.

$\Box$

It follows from Proposition 7 that if the majorization condition (\ref {cond}) is satisfied the classical capacity of the channel (\ref {REV}) is given by Corollary.

\section{Conclusion}

We have studied the mixed unitary channels generated by irreducible projective unitary representations of finite groups. 
We have calculated the capacity of such channels under some additional assumptions about the structure of the group and the probability distribution generating a mixture. It is shown that our assumptions mean that the channel can be considered as  a perturbation of quantum-classical channel of the same class. The proof is based upon the majorization theory and the theory of projective unitary representations for finite groups. We provide the text with examples illustrating the techniques for representations of Abelian as well as non-Abelian groups.

\section*{Acknowledgments} The author is grateful to A.S. Holevo, A.V. Vasilev and I.Yu. Zhdanovskii for many fruitful discussions.

\end{document}